\documentclass[12pt,a4paper]{article}

\usepackage{amsmath,latexsym,amssymb,axodraw,graphicx,cite,slashed}

\newcommand{\be}{\begin{equation}}
\newcommand{\ee}{\end{equation}}
\newcommand{\bea}{\begin{eqnarray}}
\newcommand{\eea}{\end{eqnarray}}
\newcommand{\nn}{\nonumber}

\newcommand{\dke}{\int\!\!\!\frac{d^{4-2\epsilon}k}{(2 \pi)^{4-2\epsilon}}}
\newcommand{\orbi}{{S^{1}/Z_{2}}}
\newcommand{\orbib}{{S^{1}/(Z_{2} \times Z'_{2})}}
\newcommand{\lb}{\left(}
\newcommand{\rb}{\right)}
\newcommand{\cN}{\mathcal{N}}
\newcommand{\cO}{\mathcal{O}}
\newcommand{\cL}{\mathcal{L}}
\newcommand{\bZt}{Z_{2}}
\newcommand{\sR}{\mathsf{R}}
\newcommand{\sG}{\mathsf{G}}
\newcommand{\sH}{\mathsf{H}}
\newcommand{\ds}{\slashed{\partial}}

\newcommand{\e}{\epsilon}
\newcommand{\g}{\gamma}
\newcommand{\de}{\delta}

\begin{document}

\thispagestyle{empty}
\vspace*{.5cm}
\noindent
HD-THEP-08-32 \hfill 22 December 2008

\hfill revised 14 August 2009

\vspace*{1.9cm}

\begin{center}
{\Large\bf A Realistic Unified Gauge Coupling from the\\[.4cm]
 Micro-Landscape of Orbifold GUTs}
\\[2.3cm]
{\large Christian~Gross$\,^a$ and Arthur~Hebecker$\,^b$}\\[.5cm]
{\it $^a$II. Institut f\"ur Theoretische Physik, Universit\"at Hamburg, 
Luruper Chaussee 149,\\ D-22761 Hamburg, Germany}\\[.2cm]
{\it $^b$Institut f\"ur Theoretische Physik, Universit\"at Heidelberg, 
Philosophenweg 19,\\ D-69120 Heidelberg, Germany}
\\[.4cm]
{\small\tt (\,christian.gross@desy.de\,,\,
a.hebecker@thphys.uni-heidelberg.de\,)}
\\[1.3cm]
{\bf Abstract}\end{center}
We consider 5-dimensional supersymmetric field theories where 
supersymmetry is broken by the Scherk-Schwarz mechanism (or, equivalently, 
by the $F$-term VEV of the radion). In such models, the radion effective
potential is calculable in terms of the 5d gauge coupling, the UV~cutoff 
of the 5d field theory, and the field content. We provide 
simple, explicit formulae for the leading part of the two-loop
effective potential. Our analysis applies in particular to 5d orbifold 
GUTs motivated by heterotic orbifold models. We focus on potentially 
realistic models of this type and make the additional assumption that 
the UV~cutoff scale is identical with the strong-coupling scale of the 5d 
gauge theory. Given our stabilization mechanism, the 5d radius is now
fixed in terms of the 5d gauge coupling and the field content of the model. 
This implies a prediction for the effective 4d gauge coupling only in terms
of the field content of the model. Given the `micro-landscape' provided by 
the different possible distributions of Standard Model fields between bulk 
and branes, we find a subset of models with a realistic unified gauge 
coupling. We also discuss two possibilities for the `uplifting' 
of our SUSY-breaking AdS vacua: One is based on the possible presence of 
a weak warping, the other appeals to $F$-terms in an extra brane-localized 
SUSY-breaking sector.

\newpage

\setcounter{page}{1}

\section{Introduction}

Supersymmetric grand unification at a scale of $\sim10^{16}$ GeV is one of 
the best motivated proposals for physics beyond the standard model~\cite{
Georgi:1974sy,Dimopoulos:1981yj}. It fits rather naturally into the 
framework of heterotic string theory, where a large class of potentially 
realistic constructions with gauge coupling unification can be obtained in 
orbifold model building~\cite{Buchmuller:2005jr} (see~\cite{Nilles:2008gq} 
for a recent review). One of the possibilities for overcoming the 
string-scale GUT-scale problem~\cite{Kaplunovsky:1985yy}, which generically 
affects these scenarios, is the compactification on anisotropic 
orbifolds~\cite{Ibanez:1992hc,Witten:1996mz,Hebecker:2004ce,Dundee:2008ts}, 
where at least one of the compactification radii is much larger than the 
string length scale. Such models have a useful effective description in 
terms of higher-dimensional field theories valid at energies between the 
string scale and the compactification scale. They are also known as orbifold 
GUTs which provide, independently of a possible string-theoretic UV 
completion, some of the simplest realistic models of grand 
unification~\cite{Kawamura:2000ev,Hebecker:2001wq,Asaka:2001eh}. 

It is therefore essential to understand possible stabilization mechanisms 
for the largest compact dimensions at a quantitative level. Here, we focus 
on the simple case of 5d supersymmetric gauge theories on $S^1/Z_2$ or 
$S^1/(Z_2\times Z_2')$ with hypermultiplets in the bulk and chiral matter 
localized at the boundaries. In orbifold GUTs of this type, the 4d gauge 
coupling is given by 
\be
g_4^2=\frac{g^2}{2\pi R}\,,\label{defg4}
\ee
where $g$ is the 5d gauge coupling and $R$ is the compactification radius.\footnote{We
view the 4d theory as resulting from a projection applied to the 4d spectrum 
of an \mbox{$S^1$-compactified} 5d theory. Hence the $S^1$~volume $2\pi R$ rather than
the orbifold volume $\pi R$ or $\pi R/2$ appears.
} 
It will be instructive to rewrite this relation in terms of the parameters 
\be
\frac{g_4^2N}{16\pi^2}\qquad\mbox{and}\qquad
\frac{g^2N}{24\pi^3}\equiv\frac{1}{M}\,,\label{defep}
\ee
which govern the perturbative series of an $SU(N)$ gauge theory in 4 and 
in 5 dimensions~\cite{Chacko:1999hg}.\footnote{It
has recently been shown~\cite{Panico:2006em} (footnote 4) that
considering specifically the vacuum polarization and using Pauli-Villars
regularization, the 5d loop suppression factor is actually $24 \pi^2$
rather than $24 \pi^3$. This may not be completely unexpected since in
odd dimensions, after the usual rewriting $2 \, |k|^{d-1} d|k| = (k^2)^{d/2-1} d k^2$,
 there appears a root in the integrand of the loop integral,
leading to a factor of $\pi$. We continue to use the standard
loop suppression factor, keeping in mind that a more detailed analysis may
be necessary if one wants to fix the strong-coupling scale $M$ more precisely.
}
Note that in the 5d case, loop 
corrections are proportional to positive powers of $(\Lambda/M)$, where 
$\Lambda$ is the cutoff scale. Hence $M$ can also be viewed as the 
`fundamental scale' or `strong-coupling' scale of the 5d theory: It is the 
highest scale to which the cutoff can be raised in perturbative effective 
field theory. 

In terms of the proper expansion parameters given in Eq.~(\ref{defep}), 
the expression for the 4d gauge coupling, Eq.~(\ref{defg4}), takes the form
\be
\frac{g_4^2N}{16\pi^2}=\frac{3}{4}\,\frac{1}{MR}\,.\label{defg42}
\ee
This formulation shows a rather precise connection between 4d and 5d 
perturbativity: A strongly coupled 4d effective theory emerges when 
the compactification scale is raised to the 5d strong-coupling scale $M$. 
Hence, when it comes to numbers, it is more convenient to think of $1/M$ 
rather than of $g^2$ as of the parameter defining the 5d gauge theory. 

For the phenomenological value $\alpha_{\rm GUT}\simeq 1/25$ and an $SU(5)$ 
gauge group, the l.h. side of Eq.~(\ref{defg42}) takes a value $\simeq 1/60$. 
Thus, we need a corresponding hierarchy between $1/R$ and $M$. We will see 
that such a mild hierarchy\footnote{
Independently 
of the specific value of $\alpha_{\rm GUT}$, the requirement 
of small but not extremely small 4d gauge couplings after compactification 
is common to many models with extra dimensions. Hence our analysis is 
relevant not only for higher-dimensional GUT models, but also for models 
with intermediate or TeV scale extra dimensions. 
}
is relatively easy to achieve: Given the discrete set of models provided by 
different distributions of matter fields between bulk and branes, one finds 
many situations where the Casimir energy stabilizes the radius at the 
desired scale. 

Before describing our specific results in more detail, we recall the 
generic situation: It is well-known that the compactification radius $R$ 
(i.e.~the radion field in 4d language) is a modulus at tree-level. Loop 
corrections lift the flatness of its effective potential~\cite{ 
Appelquist:1983vs}. If all bulk fields are massless, this `Casimir energy' is 
$\propto 1/R^4$ at one-loop order on dimensional grounds. Radius stabilization 
requires a more complicated functional form of the effective potential and 
hence the presence of another mass scale. This scale can be provided, for 
example, by warping~\cite{Garriga:2000jb}, by massive bulk matter or by 
brane-localized kinetic terms for bulk fields~\cite{Ponton:2001hq}. These 
and other mechanisms for radion stabilization have also been discussed by 
many authors in the context of models with spontaneously broken supersymmetry 
(see e.g.~\cite{vonGersdorff:2003rq,Luty:2002hj,Dudas:2005vn,Rattazzi:2003rj,
Correia:2006vf}). In the present, orbifold-GUT motivated context, Casimir 
stabilization has recently been analyzed in 6d, using brane-localized soft 
terms and FI-terms to provide the required mass scale~\cite{Buchmuller:2008cf}.

We base our analysis on the observation that Casimir stabilization can occur
even in the minimal realistic setting of a 5d gauge theory~\cite{
vonGersdorff:2005ce}. If it does, one has more predictive power than in many 
of the more elaborate constructions mentioned above. The idea is simply to 
use the two-loop effective potential, which is of the form $1/R^4+g^2/R^5$ for 
an $S^1$ compactification. For appropriate numerical coefficients, a \
perturbatively controlled minimum at relatively large $R$ can arise.\footnote{
Note 
that different two-loop Casimir stabilization mechanisms have been discussed 
in the context of 6d $ \lambda \phi^3$ theory~\cite{Albrecht:2001cp} and 5d 
$\lambda \phi^4$ theory~\cite{DaRold:2003yi}.
} 
For an $S^1/Z_2$ or $S^1/(Z_2\times Z_2')$ orbifold, the two-loop 
contribution is enhanced by a factor $\ln(\Lambda R)$, where $\Lambda$ is 
the UV~cutoff scale of the 5d field theory. This enhancement originates in 
the UV divergence of brane localized gauge-kinetic terms. For $\Lambda\gg 
1/R$, the logarithm is large and predictivity is maintained even in the 
presence of unknown tree-level brane operators (as long as they are not 
unnaturally large). In~\cite{vonGersdorff:2005ce}, these ideas were worked 
out in the case of $S^1$ for supersymmetric and non-supersymmetric models \
and in the case of $S^1/Z_2$, but without supersymmetry or gauge symmetry 
breaking by orbifolding. 

If we make the assumption that the cutoff or UV-completion scale $\Lambda$ 
takes its highest possible value, $\Lambda\simeq M$, the potential takes 
the form 
\be 
V(R)\quad\sim\quad \frac{1}{R^4}+\frac{g^2}{R^5}\ln(MR) \quad\sim\quad 
\frac{1}{R^4}\left(1+\frac{\ln(MR)}{MR}\right)\,.\label{genv}
\ee
The numerical coefficients of the two competing terms, which have been
suppressed for brevity, can have different signs and values. Their ratio, 
which depends only on the field content of the model, determines the 
position of the minimum. For appropriate field content, the minimum is at 
$R\gg M^{-1}$, rendering our analysis self-consistent. 

It is clear that the value of $R$ at the minimum is proportional to 
$g^2$ or, equivalently, to $1/M$. The proportionality factor is 
calculable in terms of the field content of the model. Hence 
Eq.~(\ref{defg4}) provides a prediction of the 4d gauge coupling,
even though we cannot determine the values of $M$ and $R$ independently.
Of course, there are good reasons to believe that $R^{-1}$ is of the 
order of $M_{\rm GUT}\sim 10^{16}$ GeV, which would require the 
5d model to be characterized by $M\simeq 45 M_{\rm GUT}$. However, we 
emphasize again that the overall uncertainties of these scales do not affect 
our prediction of $g_4$. This prediction is based only on the quantity 
$MR$, which is calculable in terms of the gauge group, symmetry breaking 
pattern and matter content of the 5d orbifold model. 

In the present paper, we analyze two-loop Casimir stabilization in the 
potentially realistic case of supersymmetric $ \orbi $ or $S^1/(Z_2\times 
Z_2')$ orbifolds with gauge symmetry breaking by boundary conditions. Although 
both supersymmetry (with Scherk-Schwarz breaking~\cite{Scherk:1978ta}) and 
gauge symmetry breaking have a significant effect on the Casimir energy, the 
potential can be derived essentially without new loop calculations. This 
is achieved using simple arguments based on the ${\cal N}=2$ SUSY case and 
elementary group theory. 

It is essential for our analysis that SUSY breaking is dominated 
by the $F$-term VEV of the radion superfield $T$, which contains $R$ as the 
real part of its scalar component. This situation corresponds to 
Scherk-Schwarz breaking in the rigid SUSY approximation~\cite{Marti:2001iw,
Kaplan:2001cg}. The SUSY breaking scale is proportional to the 
(dimensionless) Scherk-Schwarz twist parameter or, equivalently, $F_T$. 
This is a small number, the square of which enters the radion potential 
as an overall prefactor. Hence, the precise scale of SUSY breaking is 
irrelevant for the position of the minimum. 

We apply our general results to some $SU(5)$ orbifold GUT models. We find that 
the possibility of Casimir stabilization depends crucially on the distribution 
of matter fields between bulk and branes. (This freedom corresponds to what 
we previously called the `field content' of the model.) As described in more 
detail above, the stabilization radius in units of $g^2$ directly determines 
the value of the 4d gauge coupling at the compactification scale. Since we 
aim at potentially realistic models, we need to ensure that this value is 
consistent with the phenomenological value of the unified gauge coupling. 
Within the `micro-landscape' arising from the possible localization of matter 
fields, several examples with a realistic gauge coupling can be found. 

Our paper is organized as follows. In Sect.~2, we review the most important 
results of Ref.~\cite{vonGersdorff:2005ce} and explain our strategy for 
determining the unified gauge coupling in more detail. In Sect.~3, we derive 
general formulae for the radion potential for supersymmetric gauge theories 
on $\orbi$ with charged hypermultiplets in the bulk. This analysis is 
extended to situations with broken gauge symmetry in Sect.~4 and to 
$\orbib$ compactifications with gauge symmetry broken at one of the branes 
in Sect.~5. We apply these results to simple realistic GUT models in 
Sect.~6 and identify models in which a realistic 4d gauge coupling is 
dynamically realized. Since the Casimir energy at the stabilization point 
is negative, some form of `uplifting' is required. This issue is addressed in 
Sect.~7. The conclusions, given in Sect.~8, are followed by an Appendix, where
we describe part of the underlying component field calculation in more 
detail. Although our results, as emphasized above, can be obtained without 
any explicit new loop calculations, we find this useful in view of a 
disagreement with some of the component-field results of~\cite{Cheng:2002iz}.

\section{Perturbatively controlled radius stabilization by Casimir energy}

Let us first consider 5d gravity (with vanishing cosmological constant) and 
pure gauge theory, compactified on $S^1$:\pagebreak[0]
\begin{equation}
\int d^{4}x \int_{-\pi R}^{\pi R} \!\!\! dy \sqrt{- \det (g_{M N}) } \left( \frac{1}{2}
M_{P,5}^{3}\mathcal{R}_{5} + \frac{1}{2 g^2} \mathrm{tr} \left( F_{MN} 
F^{MN}\right) \right) \,.
\end{equation}
The parameters $M_{P,5}$ and $g$ are defined in the uncompactified 5d 
effective theory at zero momentum. We view this as an effective 
quantum field theory in which the cutoff can be taken as high as the 
strong-interaction scale. Compactifying this theory on $S^1$ then 
corresponds to an IR modification which should not introduce any new 
infinities and hence no cutoff dependence. Thus, the 4d effective potential 
for $R$ (i.e.~the Casimir energy) can only depend on $g$ and $M_{P,5}$. 
Gauge loop effects are suppressed by powers of $g^2/R$, while gravitational 
loop effects are suppressed by powers of $1/(M_{P,5} R)^3$. Hence, the 
latter are subdominant in situations where $1/g^2 \lesssim M_{P,5}
(M_{P,5}R)^2$. Neglecting gravitational interactions, we have
\begin{equation} \label{eq:V4effs1}
V(R)=\frac{1}{R^{4}}\left( c^{(1)}+c^{(2)}\frac{g^2}{R}+c^{(3)}\left( 
\frac{g^2}{R}\right) ^{2}+\ldots\right)\,,
\end{equation}
where $c^{(i)}$ is the coefficient of the $i$th loop order 
contribution.\footnote{
For
simplicity, we work with the effective potential in the Brans-Dicke frame, 
i.e., we do not absorb the prefactor $R$ of the Einstein-Hilbert term into the 
metric. Since we will eventually only be interested in models with vanishing 
cosmological constant, this will not affect the position of the minimum.
}
As explained above, the $c^{(i)}$s are cutoff-independent calculable numbers. 

Given that $c^{(1)}<0$ and $c^{(2)}>0$, one finds a minimum at 
\begin{equation}
R_{min}=-\frac{5}{4}\frac{c^{(2)}}{c^{(1)}} g^2
\end{equation}
at two-loop order.\footnote{The Casimir energy
\be
V(R_{min})=\frac{c^{(1)}}{5 R_{min}^4} \nn
\ee
is negative. In non-supersymmetric theories, one can simply introduce an 
appropriate brane tension (for models with branes or boundaries) in order 
to get a vanishing vacuum energy. In supersymmetric theories, this is less 
obvious, especially if one is not willing to compromise radion mediation as 
the dominant SUSY breaking mechanism. We will discuss this in detail in 
Sect.~\ref{sec:uplifting}.
}
Unfortunately, if all $c^{(i)}$s are $\cO(1)$, the loop expansion parameter 
$g^2/R$ is also $\cO(1)$ in the vicinity of $R\sim R_{min}$. Hence, this 
minimum is not perturbatively controlled.

However, in special cases where $c^{(1)}$ is negative and $\cO(1)$ while 
$c^{(2)}$ is large and positive, the loop expansion factor $g^2/R$ is small 
for $R$ close to $R_{min}$. Higher-loop contributions to the effective 
potential are then suppressed by powers of $g^2/R$. Perturbative control is 
guaranteed if the possible growth of $c^{(i)}\ (\textrm{for }i>2)$ with $i$
is overwhelmed by the increasing powers of $g^2/R$. This is indeed easily 
realized in simple models~\cite{vonGersdorff:2005ce}.

In essence, this strategy of finding a perturbatively controlled minimum 
by tuning the coefficients $c^{(1)}$ and $c^{(2)}$ via the field content 
also applies to orbifold compactifications. However, the power-law behavior 
of the effective potential, Eq.~(\ref{eq:V4effs1}), is in general modified. 
The reason is the presence of brane-localized operators in the effective 
action. The prime example is a brane-localized contribution to the 
gauge-kinetic term. Such terms were first studied in the context of orbifold 
GUTs, where they can be employed to achieve gauge coupling unification by a 
modified logarithmic running above the compactification
scale~\cite{Hall:2001pg,Nomura:2001mf,Hebecker:2001wq,Contino:2001si}. These
terms are logarithmically UV-sensitive. Unless the UV-completion of our
model is known, the coefficients at the cutoff scale $\Lambda$ are 
free parameters. For reasons of naturalness, we assume these to take 
$ \cO(1)$ values. However, their values at the scale $M_c = 1 / R$ are 
enhanced by an additive contribution $\sim\ln(\Lambda/M_c)$, the coefficient 
of which is calculable in the low-energy effective field theory. Thus, the 
calculability of the Casimir energy is unspoiled as long as $\Lambda$ and 
$M_c$ are at least a few orders of magnitude apart.

The log-enhanced brane-operators discussed above affect the Kaluza-Klein mass 
spectrum, resulting in an extra contribution to the one-loop Casimir energy. 
This contribution is enhanced by $\ln(\Lambda R)$ because of the running of 
the brane-operators but suppressed by $g^2/R$ because it is a brane-effect. 
Alternatively, this can be viewed as a two-loop effect since it arises in 
the interplay of the one-loop running and the one-loop Casimir energy 
calculation. This point of view is also consistent with the fact that this 
contribution is proportional to $g^{2}$. Due to the log-enhancement it 
dominates over the two-loop Casimir energy from the bulk (which is 
$\propto g^{2 }/R^5$). In summary, the 4d effective potential at leading 
two-loop order has the form
\be \label{eq:V(R)_orbi}
V(R)\,=\,\frac{1}{R^{4}}\left( c^{(1)}+c^{(br)}\,\ln(\Lambda R)\,
\frac{g^{2}}{R}
\right)\,=\,\frac{1}{R^{4}}\left( c^{(1)}+\tilde{c}^{(br)}\,\frac{\ln(
\Lambda R)}{MR}\right)
\,,
\ee
where $c^{(br)}$ (or, equivalently, $\tilde{c}^{(br)}$) is the coefficient 
of the brane-induced contribution. This coefficient will be determined for 
supersymmetric gauge theories in sections~3, 4 and 5.

As already discussed in the Introduction, we will assume that the cutoff 
scale $\Lambda$ takes its highest possible value -- the strong-coupling 
scale $M$ of the 5d gauge theory (cf.~Eq.~(\ref{defep})). 
Equation~(\ref{eq:V(R)_orbi}) then determines the stabilization radius in 
terms of $M$ and the calculable ratio $c^{(1)}/\tilde{c}^{(br)}$. 
This is the basis of our explicit determination of the unified gauge 
coupling. 

Obviously, $\Lambda$ is in principle an independent parameter of our 5d
effective theory. For example, in an orbifold compactification of the 
heterotic string, $\Lambda$ would depend on the values at which the dilaton 
and the 5 remaining compact dimensions are stabilized. As a further 
constraint, we would have to require that the 4d Planck mass is correctly 
reproduced. As discussed in some detail in~\cite{Hebecker:2004ce}, the 
present setting with a relatively large 5th dimension and a maximally 
extended validity range of the 5d gauge theory is one of the more appealing 
options for solving this complicated problem. This may be viewed as an extra 
motivation for our assumption $\Lambda\simeq M$. 

Even if $\Lambda<M$, the main message of the present analysis remains 
unchanged: Equation~(\ref{eq:V(R)_orbi}) will determine the compactification 
radius in terms of $M$, the field content, and $\Lambda$. The `micro 
landscape' of orbifold GUTs will then allow us to tune the field content in 
such a way that a realistic 4d gauge coupling is obtained. In fact, 
$\Lambda$ enters the Casimir energy only logarithmically and hence 
$g_4^2$ will also only have an (approximately) logarithmic dependence 
on $\Lambda$. Of course, our analysis breaks down if $\Lambda$ is 
so small (i.e.~the validity range of the 5d theory is so limited) that 
unknown ${\cal O}(1)$ terms are of the same size as $\ln(\Lambda/M_c)$.

\section{Casimir energy for $\orbi$} \label{sec:unbrokenG}

Before determining the Casimir energy let us briefly review 5d $\mathcal{N}$=1
SUSY and its breaking by orbifold boundary conditions.

The 5d vector multiplet (VMP) consists of a real vector $A_{M}$, a real scalar
$\Sigma$ and a Dirac spinor $\lambda$, corresponding to two 4d Weyl spinors
$\lambda_L,\lambda_R$. Under 4d $\mathcal{N}$=1 SUSY, it decomposes into a 4d
vector multiplet $ V=(A_{\mu},\lambda_L) $ and a 4d chiral multiplet $\chi =
({\Sigma+i A_5},\lambda_R) $. The `gauginos' $\lambda_L,\lambda_R$ can also be
written as an $SU(2)_R$ doublet of symplectic Majorana spinors which makes the
$SU(2)_R$ symmetry of the theory manifest~\cite{Mirabelli:1997aj}. The 5d
hypermultiplet (HMP) consists of an $SU(2)_R$ doublet of scalars $H^1,H^2$ and 
a Dirac spinor $\psi$ (which is equivalent to two Weyl spinors $\psi_L$ and
$\psi_R$).  Under 4d $\mathcal{N}$=1 SUSY, it decomposes into a 4d chiral
multiplet $H=( H^1,\psi_L) $ and another chiral multiplet $H^c=((H^2)^*,\psi_R)
$ in the conjugate representation of the gauge group.

The $\bZt$-parities of the fields can only be assigned consistently in a way
that  breaks 4d $\cN=2$ SUSY to $\cN=1$: Invariance of the action under $ \bZt$
transformations demands $ V $ to be $ \bZt$-even and $\chi$ to be $\bZt$-odd, 
while $H $ and $H^c$ must have opposite parities. Hence, only $ V $ and either 
$H $ or $H^c$ have Kaluza-Klein (KK) zero modes. The massive KK modes of the 5d
VMP at each KK level form a 4d $\cN=1$ massive vector multiplet (which has 
twice as many degrees of freedom (d.o.f.) as a massless 4d vector multiplet in
Wess-Zumino gauge).  On the other hand, the massive KK modes of $ H $ and $ H^c
$ form pairs of massive 4d chiral multiplets.

The residual SUSY can be broken by a Scherk-Schwarz twist. It has been shown
that this leads to the same spectrum as in radion mediated SUSY
breaking~\cite{Chacko:2000fn,Marti:2001iw,Kaplan:2001cg}. Hence the latter
scenario can be viewed as a dynamical realization of Scherk-Schwarz breaking.
The bosons $A_M,\Sigma$ in the VMP and the fermions $\psi_L,\psi_R$ in the HMP
are $SU(2)_{R}$ singlets and hence have `untwisted' boundary conditions. On the
other hand, the gauginos and hyperscalars are $SU(2)_{R}$ doublets which have a
nontrivial twist-matrix $T=\mathrm{exp}(2 \pi i \omega \sigma_2)$. Here,
$\omega$ is the Scherk-Schwarz parameter (which can be identified with the
radion $F$-term VEV) and $\sigma_2$ is the second Pauli matrix. As a
consequence, the KK masses of the gauginos and hyperscalars receive a shift
$n/R\rightarrow(n+\omega)/R$, which lifts the mass degeneracy of the 4d $\cN=1$
SUSY multiplets (see Fig.~\ref{fig:KKmodes}). Even though the masses of bosons
and fermions do not agree at any KK level, the UV-divergent part of the quantum
corrections respects 4d $\cN=1$ SUSY, which is \textit{locally} unbroken. In
other words, Scherk-Schwarz breaking is a global (=IR) effect which does not
modify UV properties. In particular, the logarithmic divergences at the
boundaries are supersymmetric, so that the resulting log-enhanced corrections 
to the KK masses are the same for all component fields within a 4d $\cN=1$
supermultiplet. 
\begin{figure}[ht]
\renewcommand{\arraystretch}{1.2}
{\scriptsize
\begin{tabular}{l||*{4}{p{1.2cm}}|*{4}{p{1.2cm}}}
\vdots&\vdots&\vdots&\vdots&\vdots&\vdots&\vdots&\vdots&\vdots
\\
n=3&\hrulefill&\hrulefill$\uparrow$\hrulefill&\hrulefill&\hrulefill$\uparrow$\hrulefill&\hrulefill$\uparrow$\hrulefill&\hrulefill&\hrulefill$\uparrow$\hrulefill&\hrulefill
\\
n=2&\hrulefill&\hrulefill$\uparrow$\hrulefill&\hrulefill&\hrulefill$\uparrow$\hrulefill&\hrulefill$\uparrow$\hrulefill&\hrulefill&\hrulefill$\uparrow$\hrulefill&\hrulefill
\\
n=1&\hrulefill&\hrulefill$\uparrow$\hrulefill&\hrulefill&\hrulefill$\uparrow$\hrulefill&\hrulefill$\uparrow$\hrulefill&\hrulefill&\hrulefill$\uparrow$\hrulefill&\hrulefill
\\
n=0&\hrulefill&\hrulefill$\uparrow$\hrulefill& & &\hrulefill$\uparrow$\hrulefill&\hrulefill& &
\\
\hline
KK-\#&$A_{\mu}$&$\lambda_{L}$&\mbox{$\Sigma\!+\!i A_5$}&$\lambda_{R}$&$H^{1}$&$\psi_{L}$&$(H^{2})^*$&$\psi_{R}$
\end{tabular}
}
\caption{\label{fig:KKmodes} KK masses of the components of the VMP and HMP. 
The arrows denote the states whose masses are shifted for $\omega \neq 0$.}
\end{figure}

To display the generic formula for the one-loop Casimir energy, we consider 
an $SU(2)_{R}$ doublet of complex scalars with opposite $Z_2$ 
parities on $\orbi$. Such a doublet forms the bosonic part of a 
hypermultiplet. The KK spectrum in the presence of a Scherk-Schwarz parameter
$\omega$ (allowing also for $\omega=0$) is $ m_n^{(\omega)} \equiv
(n+\omega)/R$. This gives a Casimir energy
\be
\frac 1 2 \times 4 \times \lim_{d\to 4} \frac{1}{2} \sum_{n=- \infty}^{\infty}
\int\frac{d^{d}k_{E}}{(2 \pi)^{d}}\ln\left(k_{E}^2+ (m^{(\omega)}_n)^{2} 
\right)\,,
\ee
where the factor $ 1/2 $ comes from the $Z_2$ projection, the factor 
$ 4 $ counts the d.o.f. of two complex scalars, and $k_{E}$ is a 
Euclidean $d$-momentum. The finite, $ R $-dependent part of the above 
expression is\footnote{The one-loop vacuum diagrams can be split into an 
$R$-dependent finite part and a divergent part which is linear in $R$ and 
represents a contribution to the 5d cosmological constant~\cite{
Appelquist:1983vs} (see also~\cite{vonGersdorff:2008df}). The latter cancels 
between bosons and fermions in supersymmetric models.}~\cite{Antoniadis:1998sd}
\be
4 \, f(\omega , R)\equiv -\frac{6 \, \zeta_{\omega}(5)}{(2 \pi)^{6} R^4} \, ,
\ee
where $\zeta_{\omega}(s) \equiv \sum_{k=1}^{\infty} k^{-s}\cos(2\pi k\omega)$
is a generalization of the Riemann zeta-function~$ \zeta_{\omega=0}$. In a
theory with Scherk-Schwarz SUSY breaking, each d.o.f. with $ \omega=0 $ has a
superpartner with non-zero $ \omega $, so that the following quantity is 
useful:
\be
c_{\omega}\equiv 2 R^4 \left( f(\omega , R)-f(0, R)\right) = \frac{3}{(2
\pi)^{6}} \left( \zeta(5)-\zeta_{\omega}(5)\right) \,.
\ee
Note that $ c_{\omega}\ge 0$, since $\zeta(s)>\zeta_{\omega}(s)$ for $\omega
\neq 0$.

\subsection*{Casimir energy for a pure gauge theory}
\label{sec:puregaugetheory}

The one-loop bulk coefficient for a pure gauge theory with gauge group $ G $ 
and supergravity is obtained by adding the contributions from all physical 
d.o.f., taking into account a minus sign for fermions and the respective 
Scherk-Schwarz parameter $ \omega $ of each field. One finds\footnote{
Let 
us fix here our group theory conventions. The generators $ T_{\sR}^{a} $ for 
an irreducible representation $ \sR $ are normalized such that $\textrm{tr}
(T_\sR^{a}T_\sR^{b})=C_\sR \delta^{a,b}$, where $ C_\sR $ is the Dynkin index. 
The quadratic Casimir operator is denoted by $ C_2(\sR) $. For the dimension 
of the representation $ \sR $ we use the notation $ d_{\sR} $. The adjoint 
representation is denoted by $ \sR= \sG $. We often use the identity $
d_{\sR}C_2(\sR)=d_{\sG}C_{\sR} $.
}~\cite{Antoniadis:1998sd,vonGersdorff:2002tj}
\be \label{eq:c^1_vmp_bulk}
 c^{(1)}= c^{(1)}_{vmp}+ c^{(1)}_{grav} \qquad \textrm{where} \qquad
c^{(1)}_{vmp}=-2 c_{\omega} d_\sG\,,  \qquad  c^{(1)}_{grav}=-4 c_{\omega} \, .
\ee

Let us now determine $ c^{(br)} $. This contribution comes from brane-localized
operators. Such operators induce a shift $ \delta m_n $ of the KK masses. The
relative mass shift $ \Delta\equiv \delta m_n/m_n  $ is independent of the KK
level~$ n $. For one bosonic d.o.f. with tree level KK spectrum $
m_n^{(\omega)}$, the extra contribution to $V(R)$ due to brane-localized
operators is given by~\cite{vonGersdorff:2005ce}
\bea \label{eq:Vbr}
V^{(br)}(R)&\equiv &\frac{1}{2}\lim_{d\to 4} \frac{1}{2} 
\sum_{n=-\infty}^{\infty}\int\frac{d^{d}k_{E}}{(2 \pi)^{d}}\left\{ \ln\left( 
k_{E}^2+(1+\Delta)^2(m^{(\omega)}_n)^{2} \right)-\ln\left( k_{E}^2+ 
(m^{(\omega)}_n)^{2}
\right)\right\} \nonumber
\\
&=& 4 \ \Delta \ f(\omega , R) \ + \cO(\Delta^2)\,.
\eea
To obtain the entire contribution to the Casimir energy, one has to sum over 
all d.o.f.

The brane-localized gauge-kinetic terms which are induced in case of a gauge
theory are equivalent to a shift of the 4d effective gauge coupling $ g_4 $
by~\cite{Hebecker:2001wq,Hall:2001pg,Nomura:2001mf,Contino:2001si,
Hebecker:2002vm}
\be \label{eq:delta_g4}
\delta(g_4^{-2})=-\frac{1}{4 \pi^{2}} C_{\sG} \ln(MR) \,.
\ee
This in turn corresponds to a relative shift of the KK masses of the VMP by
\be \label{eq:delta_m}
\Delta_{vmp} = -  \frac{g^2}{2 \pi R} \delta(g_4^{-2}) = \frac{1}{8 \pi^{3}}
C_{\sG} \ln(MR) \frac{g^2}{R}\,.
\ee
Note that, in this equation, a factor $1/2$ arising from the fact that 
$m^2$ is corrected by $2\Delta_{vmp}$ cancels a factor $2$ arising from the 
enhanced sensitivity of cosine-modes to brane terms as compared to the zero 
mode. Since $ \Delta_{vmp} $ is the same for all d.o.f. of the VMP, using
Eq.~(\ref{eq:Vbr}) and adding the contributions from all d.o.f. we find
\be \label{eq:cbr_propto_c1}
V^{(br)}_{vmp}(R)\equiv c^{(br)}_{vmp}  \ \ln(M R) \ \frac{g^{2}}{R^5} = 4 \
\Delta_{vmp} \ c^{(1)}_{vmp} \frac{1}{R^4} \, . 
\ee
Using this result and Eqs.~(\ref{eq:c^1_vmp_bulk}) and (\ref{eq:delta_m}) one
can easily read off
\be \label{eq:c1br_puregauge}
c^{(br)}_{vmp}=-\frac{c_{\omega}}{\pi^3} d_\sG  C_{\sG}\,  .
\ee
Note that this is always negative so that perturbatively controlled radius
stabilization cannot be achieved in a pure super-Yang-Mills theory.

\subsection*{Including hypermultiplets}

Including the physical d.o.f. of HMPs, the one-loop bulk coefficient
is~\cite{Antoniadis:1998sd,vonGersdorff:2002tj}
\be \label{eq:c1bulk}
c^{(1)}=c^{(1)}_{hmp}+c^{(1)}_{vmp}+c^{(1)}_{grav}=2 c_{\omega} (d_\sR -d_\sG
-2) \, .
\ee

We now determine $ c^{(br)} $ in the presence of HMPs. As mentioned before, $
c^{(br)} $ is due to brane-localized operators induced by quantum fluctuations
above the compactification scale. More precisely, $ c^{(br)}_{vmp} $ respectively $
c^{(br)}_{hmp} $ denote the contribution from the one-loop selfenergy of the 
VMP respectively HMP. First of all, notice that, for both cases, the contribution 
from the HMP in the loop vanishes. To see why, recall that $ H $ and $ H^c $ 
have opposite $ \bZt $ parities. Fields with opposite parities however lead to
brane-localized operators of opposite signs.\footnote{
The 
reason is the following~\cite{Contino:2001si}: For an $ S^1 $ compactification 
there are no 4d boundaries where logarithmic divergences can occur. As a 
consequence, logarithmic divergences due to the even KK modes have to be 
canceled by the divergences of the odd KK modes. Thus, on $\orbi$, fields 
with even and odd $Z_2$ parities give opposite log-divergences.
} 
Hence, the contributions from $ H $ and $ H^c $ cancel each other and only the
VMP in the loop contributes to the selfenergy.

This means that $ c^{(br)}_{vmp} $, as given in Eq.~(\ref{eq:c1br_puregauge}),
is unchanged. In the following we derive $ c^{(br)}_{hmp} $. To start with, let
us consider a HMP in the adjoint representation. A 5d supersymmetric gauge
theory with a VMP and an \textit{adjoint} HMP has 5d $\cN=2$ SUSY, 
corresponding to $\cN=4$ in 4d.\footnote{
This 
can for instance be verified by dimensional reduction of 10d supersymmetric 
gauge theory.
}
Now, note that one of the four 4d SUSY parameters is invariant under modding 
out by reflections and translations: Two of the SUSY parameters are even under
$\bZt$-reflections and one of these furthermore has no Scherk-Schwarz twist 
(cf. Fig.~\ref{fig:KKmodes} where $\psi_{L}$ is an invariant Weyl spinor). 
This means that after compactification on $ \orbi $ the theory still has some 
unbroken SUSY so that the vacuum energy must vanish. Thus, the contribution 
of the adjoint HMP has to cancel that of the VMP:
\be \label{eq:c1br_adjhmp}
c^{(br)}_{adj. hmp}=\frac{c_{\omega}}{\pi^3} d_\sG  C_{\sG}\, .
\ee
This corresponds to a KK mass shift $ \Delta_{adj. hmp}=\Delta_{vmp} $.

Next we generalize this to a HMP in an arbitrary irreducible
representation~$\sR$. The only quantity that can change is the `group theory
factor' $d_\sG C_{\sG}$ in Eq.~(\ref{eq:c1br_adjhmp}). Clearly, there is more
than one expression for a general $ \sR $ which in the special case $ \sR = \sG
$ reduces to Eq.~(\ref{eq:c1br_adjhmp}) (e.g. both $ C_2(\sR)$ and $C_{\sR}$
reduce to $C_{\sG}$). The correct generalization of Eq.~(\ref{eq:c1br_adjhmp})
is found by recalling that the brane-localized operators arise from the 
one-loop selfenergy of the HMP with a VMP in the loop. Since the 
corresponding `coupling matrices' are given by $(T_{\sR}^a)_{i j}$, the mass 
shift for the HMP component with gauge group index $ i $ (the result has to 
be independent of $ i $, of course, due to the unbroken gauge symmetry) is
\vspace{-15pt}
\be
\Delta_{hmp}  \propto\sum_{a,j} 
{
\unitlength=1.0 pt
\SetWidth{0.5}      
\begin{picture}(100,40)(0,0)
\put(0,0){\line(1,0){100}}
\put(50,0){\oval(40,40)[t]}
\put(5,3){$i $}
\put(95,3){$i $}
\put(50,-12){$j $}
\put(50,23){$a $}
\end{picture} 
} 
\quad
\propto  \sum_a (T_{\sR}^a T_{\sR}^a)_{ii} \equiv C_2(\sR)\, .
\ee
Hence, the correct generalization of Eq.~(\ref{eq:c1br_adjhmp}) is
\be \label{eq:c1br_hmp}
c^{(br)}_{hmp}=\frac{c_{\omega}}{\pi^3} d_{\sR} C_2(\sR) =
\frac{c_{\omega}}{\pi^3} d_\sG C_{\sR}\, .
\ee
The total contribution of both a VMP and a HMP takes the simple form
\be \label{eq:c1br_total}
\qquad c^{(br)}=\frac{c_{\omega}}{\pi^3} d_\sG \left[  C_\sR- C_\sG \right] 
\, .
\ee

As an example, consider the case $G=SU(N)$ and $ h $ HMPs in the fundamental
representation (so that $ C_{\sR}\to hC_{\mathsf{F}}=h/2 $). One can easily 
check that for any number $ h $ it is impossible to have $c^{(1)}=2 c_{\omega} 
(h N -N^2 -1)<0$ and at the same time ${c^{(br)}=c_{\omega}/\pi^3 (N^2-1)
[h/2-N]>0}$. This situation may be improved if some of the gauge symmetry is 
broken. We discuss this in the following.

\section{$ \orbi $ with gauge symmetry breaking}\label{sec:orbi-broken}

Orbifold boundary conditions may break some of the bulk gauge symmetry at the
boundaries~\cite{Kawamura:2000ev}. Let us assume a breaking $G \rightarrow H =
H_1 \otimes \ldots \otimes H_n$, where the $H_i$ are the simple factors and
$U(1)$ factors. The generators $ T^a $ of $ G $ are accordingly split into a 
set of generators $ T^{\bar a} $ of $ H $ and a set of `broken generators' 
$T^{\hat a}$. The HMP representation $ \sR $ of $ G $ splits into $\bigoplus_k 
\sR_k$ where each $ \sR_k $ is a representation of $H = H_1 \otimes \ldots 
\otimes H_n $.

The one-loop bulk coefficient, Eq.~(\ref{eq:c1bulk}), remains unchanged for $
\orbi $ compactifications with broken gauge symmetry. The reason is simply that
the KK mass spectrum is unchanged in comparison to the unbroken case (even
though the wavefunctions of the higher KK modes of some components of the VMP
and HMP have flipped $\bZt$-parity). As we will see in Sect.~\ref{sect:orbib},
this is different for $\orbib $ compactifications, where also the one-loop bulk
Casimir energy `feels' the gauge symmetry breaking.

Since at the boundaries the gauge symmetry is broken, the boundary coefficient,
Eq.~(\ref{eq:c1br_total}), is modified. Let us first consider $c^{(br)}_{vmp}$.
As argued in Sect.~\ref{sec:unbrokenG}, only the VMP in the loop gives a
contribution. $\bZt$-invariance of the action implies that the structure
constants $f^{a b c}$ have an even number of `broken
indices'~\cite{Hall:2001pg}. The brane-localized operators leading to
$c^{(br)}_{vmp}$ are thus determined from diagrams with the following gauge
group indices:
\be
{
\unitlength=1.0 pt
\SetScale{1.0}
\SetWidth{0.5}      
\begin{picture}(70,40)
\Text(-10,40)[]{1.)}
\Line(0,10)(70,10)
\CArc(35,10)(15,0,180)
\Text(5,15)[]{$\bar c$}
\Text(65,15)[]{$\bar c$}
\Text(35,33)[]{$\bar a$}
\Text(35,3)[]{$\bar b$}
\end{picture} 
\qquad
\begin{picture}(70,40)
\Text(-10,40)[]{2.)}
\Line(0,10)(70,10)
\CArc(35,10)(15,0,180)
\Text(5,15)[]{$\bar c$}
\Text(65,15)[]{$\bar c$}
\Text(35,33)[]{$\hat a$}
\Text(35,3)[]{$\hat b$}
\end{picture} 
} \nn
\ee
A crucial point is that the prefactor of the brane-localized terms induced by
diagrams of type  1.) is minus that of diagrams of type  2.). The reason is 
the same which allowed us to argue that the contribution from the HMP in the 
loop vanishes: Fields with opposite parities give contributions of opposite 
signs. Thus, the mass shift of the unbroken VMP components with index $ \bar 
c $  is (up to a prefactor which doesn't depend on group theory indices)
\be \label{eq:Delta_vmp_broken}
\Delta^{\bar c}_{vmp} \propto  +\sum_{\bar a , \bar b} f^{\bar a \bar b \bar c}
f^{\bar a \bar b \bar c} - \sum_{\hat a ,\hat b} f^{\hat a \hat b \bar c}
f^{\hat a \hat b \bar c} = \sum_{a,b}\eta^a f^{ab\bar c}  f^{ab\bar c}\,,
\ee
where we defined $ \eta^{\bar a}=1,  \eta^{\hat a}=-1$. Since the tree-level KK
masses of all d.o.f. are the same, one has
\be \label{eq:cbrvmpcalc}
c^{(br)}_{vmp}\propto \sum_{\bar c}\Delta^{\bar c}_{vmp} \propto \sum_{a, b,
c}\eta^a f^{abc} f^{abc} =(2 d_\sH- d_\sG)C_{\sG} \, ,
\ee
where we used  $\sum_{\bar a ,\hat b} f^{\bar a \hat b \hat c} f^{\bar a \hat b
\hat c} - \sum_{\hat a , \bar b}f^{\hat a \bar b \hat c} f^{\hat a \bar b \hat
c} =0 $. We thus have\footnote{In order to calculate $ c^{(br)}_{vmp} $ one
could also proceed as in Sect.~\ref{sec:puregaugetheory} and use the shifts
$\delta(g_{4,i}^{-2})$ of the 4d gauge couplings of the unbroken subgroups
$H_i$, which can be extracted from Ref.~\cite{Hebecker:2002vm}:
\be
\delta( g_{4,i}^{-2}) =-\frac{1}{4 \pi^{2}}\left(  2 C_{2}(\sH_i) - C_{2}(\sG)
\right)  \ln(MR)\, . \nn
\ee
One arrives at the result
\be \label{eq:brokencoeff}
c^{(br)}_{vmp}=-\frac{c_{\omega}}{\pi^3} \left\{ 2 \left( d_{\sH_1} C_2(\sH_1)+
\ldots + d_{\sH_n} C_2(\sH_n) \right)  -d_\sH C_2(\sG) \right\}\,. \nn
\ee
The above formula however depends on all factors $H_i$ of $ H $ and hence
conceals the fact that really the only information from the gauge symmetry
breaking entering the result is the number $ d_{\sH} $.
}
\be \label{eq:cbr_vmp_broken}
c^{(br)}_{vmp}=-\frac{c_{\omega}}{\pi^3} (2 d_\sH- d_\sG)C_{\sG} \,.
\ee

In a similar way we can infer $c^{(br)}_{hmp}$, which is due to the selfenergy
of the HMP with a VMP in the loop. Let us split the indices $\{i\}$ of the HMP 
representation into two sets $\{\bar\imath\}$ and $\{\hat\imath\}$ by defining 
that, say, $H^{\bar \imath}$ is $\bZt$-even and $H^{\hat \imath}$ is 
$\bZt$-odd. Then, only elements $(T_{\sR}^a)_{i j}$ with an even number of 
`hatted indices' $ \hat a $, $\hat\imath$ or $ \hat \jmath $ are nonvanishing 
in order for the interaction terms to be $ \bZt  $-invariant. Thus, the group 
theory factor is determined by:
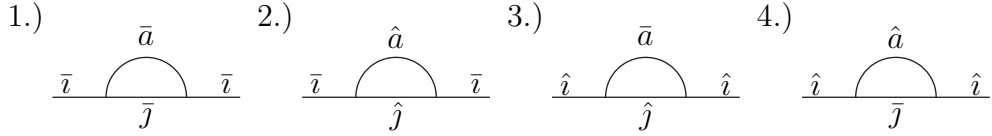
\be
{
\unitlength=1.0 pt
\SetScale{1.0}
\SetWidth{0.5}      
\begin{picture}(70,40)
\Text(-10,40)[]{1.)}
\Line(0,10)(70,10)
\CArc(35,10)(15,0,180)
\Text(5,15)[]{$\bar \imath$}
\Text(65,15)[]{$\bar \imath$}
\Text(35,33)[]{$\bar a$}
\Text(35,3)[]{$\bar \jmath$}
\end{picture} 
\qquad
\begin{picture}(70,40)
\Text(-10,40)[]{2.)}
\Line(0,10)(70,10)
\CArc(35,10)(15,0,180)
\Text(5,15)[]{$\bar \imath$}
\Text(65,15)[]{$\bar \imath$}
\Text(35,33)[]{$\hat a$}
\Text(35,3)[]{$\hat \jmath$}
\end{picture} 
\qquad
\begin{picture}(70,40)
\Text(-10,40)[]{3.)}
\Line(0,10)(70,10)
\CArc(35,10)(15,0,180)
\Text(5,15)[]{$\hat \imath$}
\Text(65,15)[]{$\hat \imath$}
\Text(35,33)[]{$\bar a$}
\Text(35,3)[]{$\hat \jmath$}
\end{picture} 
\qquad
\begin{picture}(70,40)
\Text(-10,40)[]{4.)}
\Line(0,10)(70,10)
\CArc(35,10)(15,0,180)
\Text(5,15)[]{$\hat \imath$}
\Text(65,15)[]{$\hat \imath$}
\Text(35,33)[]{$\hat a$}
\Text(35,3)[]{$\bar \jmath$}
\end{picture} 
} \nn
\ee
Note that, in contrast to VMP loops, the effect of HMP loops is non-zero
even if the external index is $Z_2$-odd (cf. diagrams 3.) and 4.)).

For the same reason as above for $c^{(br)}_{vmp}$, 1.) and 2.),
respectively 3.) and 4.), have prefactors of opposite signs. Thus we have
\bea \label{eq:delta_hmp_i}
\Delta^{\bar\imath}_{hmp} &\propto & +\sum_{\bar a ,\bar \jmath}(T_{\sR}^{\bar
a})_{\bar \imath \bar \jmath} (T_{\sR}^{\bar a})_{\bar \jmath \bar \imath} -
 \sum_{\hat a ,\hat \jmath} (T_{\sR}^{\hat a})_{\bar \imath \hat \jmath}
(T_{\sR}^{\hat a})_{\hat \jmath \bar \imath} = \sum_a \eta^a (T^a_{\sR}
T^a_{\sR})_{\bar \imath \bar \imath}  \nn 
\\
\Delta^{\hat \imath}_{hmp}  &\propto &+\sum_{\bar a ,\hat \jmath} 
(T_{\sR}^{\bar a})_{\hat \imath \hat \jmath} (T_{\sR}^{\bar a})_{\hat \jmath
\hat \imath} - 
\sum_{\hat a ,\bar \jmath} (T_{\sR}^{\hat a})_{\hat \imath \bar \jmath}
(T_{\sR}^{\hat a})_{\bar \jmath \hat \imath} = \sum_a \eta^a (T^a_{\sR}
T^a_{\sR})_{\hat \imath \hat \imath} \,.
\eea
Note that for $\eta^a=1$ $ \forall a$, both expressions reduce to $C_2(\sR)$.

In order to calculate the Casimir energy, we furthermore need to know the
(relative) prefactors which are missing in Eq.~(\ref{eq:delta_hmp_i}). To this
end recall that $ H $ and $H^c$ have opposite parities and $ H^c $ transforms
under  $ \bar \sR $ if $H$ transforms under $ \sR $. Thus, 1.) corresponds to
even modes in representation $\sR$ (from $ H^{\bar \jmath} $) and odd modes in
rep. $ \bar \sR $  (from $ H^{c \bar \jmath} $), while 3.) corresponds to odd
modes in rep. $ \sR $ (from $ H^{\hat \jmath} $) and even modes in rep. $ \bar
\sR $  (from $ H^{c \hat \jmath} $). This shows that the interchange $\bar
\jmath \leftrightarrow \hat \jmath$ corresponds to $ \sR \leftrightarrow \bar
\sR $. Since $(T_{\bar \sR}^a)_{ij}=-(T_{\sR}^a)_{ji}$, the proportionality 
factors between $ \Delta^{\bar\imath}_{hmp}$, $ \Delta^{\hat \imath}_{hmp}$ 
and the r.h. sides of Eq.~(\ref{eq:delta_hmp_i}) are the same. Thus we have
\be 
c^{(br)}_{hmp}\propto  + \sum_{\bar \imath} \Delta^{\bar \imath}_{hmp} +
\sum_{\hat \imath} \Delta^{\hat \imath}_{hmp}  \propto \sum_{a, i, j}\eta^a  
(T_{\sR}^{ a})_{i j} (T_{\sR}^{ a})_{j i}=(2 d_\sH- d_\sG)C_{\sR} \, .
\ee
The final result is then
\be
c^{(br)}_{hmp}=\frac{c_{\omega}}{\pi^3} (2 d_\sH- d_\sG)C_{\sR} \,.
\ee
Adding $ c^{(br)}_{vmp} $ and $ c^{(br)}_{hmp} $, we get the simple result
\be \label{eq:orbi-broken-result}
c^{(br)}=\frac{c_{\omega}}{\pi^3} (2 d_\sH- d_\sG) \big[ C_\sR- C_\sG \big]\, .
\ee
We stress that this does not depend on the details of the gauge symmetry
breaking, but only on the dimension of $H$.

To illustrate this result we consider again $G=SU(N)$ and $h$ fundamental HMPs.
One has $C_\sR-C_\sG\to h/2-N$ in this case. This is negative, unless $h\geq 2 
N$ which would however imply $ c^{(1)}>0 $. Hence, $(2 d_\sH -d_\sG)$ needs to 
be negative in order to obtain $c^{(br)}>0$ and $ c^{(1)}<0 $. The only 
possible breaking pattern of $SU(N)$ by $\bZt$ inner automorphisms is $SU(p+q)
\rightarrow SU(p)\times SU(q) \times U(1)$~\cite{Slansky:1981yr} (see
also~\cite{Hebecker:2001jb}), for which  $2 d_\sH -d_\sG=(p-q)^2 -1$. This is
negative only for $p=q$. A potentially phenomenologically interesting example
for this is $SU(6)\rightarrow SU(3)\times SU(3) \times
U(1)$~\cite{Burdman:2002se}. On the other hand, for the important case
$SU(5)\rightarrow SU(3)\times SU(2)\times U(1)$, the term $(2d_\sH -d_\sG)$ is
zero, i.e.~the contribution of the bulk fields to the leading order two-loop
term vanishes.

\section{$\orbib$ with broken gauge symmetry at one brane} \label{sect:orbib}

The space $\orbib$ is obtained by modding out by a second $ Z'_2 $ parity. In this case, fields can have different boundary conditions at the two inequivalent fixed points. If some of the gauge symmetry is broken at both of the boundaries, there are additional massless fields at tree level besides the radion, namely the zero modes of some of the higher-dimensional components of the gauge bosons.\footnote{This is also the case for $ \orbi $ compactification with broken gauge symmetry. 
}
They acquire masses from radiative corrections~\cite{Haba:2002py}.
In this section we restrict our attention to situations where the gauge symmetry remains unbroken at one of the branes. In that case the radion is the only modulus.

\subsection*{Pure gauge theory}

As in the previous sections we start with a pure gauge theory. The $(Z_2,Z'_2)$-parities and the KK levels of the 4d superfields $( V,\chi )$ which form the VMP are shown in the following table (cf.~\cite{Nomura:2001mf,Hebecker:2001wq,Hall:2001pg}):
\begin{center}
\begin{tabular}{|c|c|c|}
\hline
d.o.f. & $(Z_2,Z'_2)$-parity & KK spectrum \\
\hline
$V^{\bar a}$ & $(+,+)$  & $2n/R$\\
$V^{\hat a}$ & $(+,-)$ & $(2n+1)/R$ \\
$\chi^{\bar a}$ & $(-,-)$  & $(2n+2)/R$\\
$\chi^{\hat a}$ & $(-,+)$  & $(2n+1)/R$ \\
\hline
\end{tabular}
\end{center}
Note that the KK masses of broken and unbroken components of a multiplet are displaced. This is an important difference to the $ \orbi $ case, which has the consequence that also the one-loop bulk Casimir energy feels the breaking of the gauge symmetry as we will see.

The Casimir energy for one bosonic d.o.f. on $\orbib $ with KK spectrum ${ 2(n+ \omega) /R }$ is given by $ f(\omega, R/2) $ while for one d.o.f. with KK spectrum ${ [2(n+ \omega) +1]/R }$ it is ${ f(\omega +1/2, R/2) }$. Using the duplication formula ${ Li_{s}(z)+Li_{s}(-z)= 2^{1-s}Li_{s}(z^2) }$ for polylogarithms ${ Li_{s}(z) \equiv \sum_{k=1}^{\infty}k^{-s} z^k }$, one finds ${ \zeta_{\omega+ \frac 1 2}(5)=-\zeta_{\omega}(5) + 1/16 \ \zeta_{2 \omega}(5) }$. This, together with the inverse quartic scaling of $f(\omega,R)$ with $R$, results in
\bea \label{eq:fofomega}
f(\omega,R/2)&=& + 16 f(\omega,R) \nn
\\
f(\omega +1/2, R/2)&=&-16 f(\omega,R) + f(2 \omega,R) \,.
\eea
Fields with odd KK spectrum give an almost opposite contribution to the Casimir energy as fields with even KK spectrum.

By adding the contributions from all d.o.f. -- taking into account the boundary conditions and spin of each d.o.f. -- one finds with the help of Eq.~(\ref{eq:fofomega}) that
\be \label{eq:v1_vmp_orbib}
c^{(1)}_{vmp}=-32 (2 d_{\sH} - d_{\sG} ) c_{\omega} - 2 (d_{\sG} - d_{\sH}) c_{2 \omega} \,.
\ee
As a simple check, for the unbroken case $ \sH = \sG $ this becomes $ c^{(1)}_{vmp}=-32  d_{\sG}  c_{\omega}  $, which is $ 16 $ times the result for $ \orbi $. The relative factor of $ 16(=2^4) $ arises, because the length of the physical space for $\orbib$ is one half of that for $\orbi$.

We now determine the brane-coefficient. The UV-divergent contribution to the brane-localized operators is induced by fluctuations of the bulk fields `close to' the brane. It can therefore not depend on the boundary conditions at the other brane. This implies that for the unbroken brane of $\orbib$ we can use the mass shift for $\orbi$ with unbroken gauge symmetry (cf. Eq.~(\ref{eq:delta_m})), and for the broken brane we can use the mass shift for $\orbi$ with broken gauge symmetry (cf. Eq.~(\ref{eq:Delta_vmp_broken})). More precisely,  we have to take one half of Eq.~(\ref{eq:delta_m}) respectively Eq.~(\ref{eq:Delta_vmp_broken}), since we need the contribution of only one brane. The argument of this paragraph is also valid for the HMP contribution discussed below.

The gauge coupling correction, or equivalently the mass shift of the KK modes (Eq.~(\ref{eq:delta_m})), which determines the coefficient $c^{(br,G)}_{vmp}$ for the unbroken brane, is $G$-universal. Thus, Eq.~(\ref{eq:cbr_propto_c1}) applies, and together with Eq.~(\ref{eq:v1_vmp_orbib}) we immediately get
\be \label{eq:c_brG_vmp}
c^{(br,G)}_{vmp}= - \frac{8}{\pi^3} C_{\sG} \left[ (2 d_{\sH} - d_{\sG} ) c_{\omega} + (d_{\sG} - d_{\sH}) \frac{c_{2 \omega}}{16} \right] \,.
\ee

In order to determine the coefficient $c^{(br,H)}_{vmp}$ for the brane where the gauge symmetry is reduced to $H$, we use Eq.~(\ref{eq:Delta_vmp_broken}). Moreover, since no brane-localized terms are induced for the broken components of the VMP (i.e.~those with `shifted' KK spectrum $(2n+1)/R$) at that brane, the summation Eq.~(\ref{eq:cbrvmpcalc}) still applies and we get
\be  \label{eq:c_brH_vmp}
c^{(br,H)}_{vmp}= - \frac{8 \,c_{\omega}}{\pi^3} C_{\sG} (2 d_{\sH} - d_{\sG} ) \,.
\ee
The factor $8$ in comparison to Eq.~(\ref{eq:cbr_vmp_broken}) arises, since we have to multiply by $ 2^4 $ (for the reduced length) and divide by two (for \textit{one} brane).

\subsection*{Including hypermultiplets}

The $(Z_2,Z'_2)$-parities of the 4d superfields $H$ and $H^c$ which form the HMP follow from the $(Z_2,Z'_2)$-parities of the VMP.  $H$ and $H^c$ necessarily have opposite $Z_2$-parities and opposite $Z'_2$-parities, leading to a chiral spectrum for the zero modes.  The $Z'_2$-parities depend on the gauge group index. We still have the freedom to choose an overall sign of the $Z'_2$ action on the HMP. Similar to the $\orbi$-case, we define, by the following table, two sets of indices $\{\bar \imath\}$ and $\{\hat \imath\}$:
\begin{center}
\begin{tabular}{|c|c|c|}
\hline
d.o.f. & $(Z_2,Z'_2)$-parity & KK spectrum \\
\hline
$H^{\bar \imath}$ & $(+,+)$  & $2n/R$\\
$H^{\hat \imath}$ & $(+,-)$ & $(2n+1)/R$ \\
$H^{c \, \bar \imath}$ & $(-,-)$  & $(2n+2)/R$\\
$H^{c \, \hat \imath}$ & $(-,+)$  & $(2n+1)/R$ \\
\hline
\end{tabular}
\end{center}
The number of $\bar \imath$-indices is denoted by $d_1$ and the number of $\hat \imath$-indices is denoted by $d_2$, so that $d_1+d_2=d_{\sR}$. Using Eq.~(\ref{eq:fofomega}), the bulk coefficient for the HMP is found to be
\be \label{eq:v1_hmp_orbib}
c^{(1)}_{hmp}= 32 (d_1-d_2)\, c_{\omega} + 2 \, d_2 \, c_{2 \omega} \,.
\ee

The brane-coefficient from the unbroken brane, denoted by $c^{(br,G)}_{hmp}$, follows from the results of Sect.~\ref{sec:unbrokenG}: From Eqs.~(\ref{eq:c1bulk}) and (\ref{eq:c1br_hmp}), together with Eq.~(\ref{eq:cbr_propto_c1}), one reads off that (for one brane)
\be \label{eq:delta_m_hmp}
\Delta_{hmp} = \frac{1}{16 \pi^{3}} C_2(\sR) \ln(MR) \frac{g^2}{R}\,.
\ee
This and Eq.~(\ref{eq:cbr_propto_c1}) lead to
\be \label{eq:c_brG_hmp}
c^{(br,G)}_{hmp}=  \frac{8}{\pi^3} C_2(\sR) \left[ ( d_1-d_2) c_{\omega} + d_2 \frac{c_{2 \omega}}{16}   \right] \,.
\ee
Regarding the contribution $c^{(br,H)}_{hmp}$, Eq.~(\ref{eq:delta_hmp_i}) together with the subsequent paragraph implies (for $i=\bar \imath$ and $i=\hat \imath$)
\be \label{eq:deltam_hmp-broken}
\Delta^i_{hmp} = \frac{1}{16 \pi^{3}} \sum_a \eta^a (T^a_{\sR}T^a_{\sR})_{ii} \ln(MR) \frac{g^2}{R}\,.
\ee
Using this and Eq.~(\ref{eq:Vbr}), one arrives at
\be  \label{eq:c_brH_hmp}
c^{(br,H)}_{hmp}=  \frac{8}{\pi^3} \left[\left( \sum_{a, \bar \imath} \eta^a (T^a_{\sR}T^a_{\sR})_{\bar \imath \bar \imath} - \sum_{a, \hat \imath} \eta^a (T^a_{\sR}T^a_{\sR})_{\hat \imath \hat \imath}\right) c_{\omega} + \left( \sum_{a, \hat \imath} \eta^a (T^a_{\sR}T^a_{\sR})_{\hat \imath \hat \imath}\right)  \frac{c_{2 \omega}}{16} \right] \,.
\ee
As a simple check, one can verify that for an adjoint HMP, the expression $\sum_{a, \hat \imath} \eta^a (T^a_{\sR}T^a_{\sR})_{\hat \imath \hat \imath}$ vanishes, so that in this case one indeed has $c^{(br,H)}_{hmp}=-c^{(br,H)}_{vmp}$.

The above results (\ref{eq:v1_hmp_orbib}), (\ref{eq:c_brG_hmp}) and (\ref{eq:c_brH_hmp}) depend on $d_1$ and $d_2$. This dependence disappears if one has a second HMP with opposite $Z'_2$-parities, but the same quantum numbers. The flip of the $Z'_2$-parities corresponds to an interchange of $d_1$ and $d_2$. Then Eqs.~(\ref{eq:v1_hmp_orbib}), (\ref{eq:c_brG_hmp}) and (\ref{eq:c_brH_hmp}), for the combined effect of such a pair of HMPs, simplify to
\bea \label{eq:hmp-double}
c^{(1)}_{hmp+hmp'} &=&  2 \, d_{\sR} \, c_{2 \omega} \nn
\\
c^{(br,G)}_{hmp+hmp'}&=&  \frac{1}{2 \pi^3} d_{\sG} \,C_{\sR}\, c_{2 \omega} \nn
\\
c^{(br,H)}_{hmp+hmp'}&=&  \frac{1}{2 \pi^3} (2 d_{\sH}-d_{\sG})\, C_{\sR} \,c_{2 \omega} \,.
\eea

\section{Application to 5d SUSY-GUT models}

For the application of our results to realistic models, we need to consider also the effect of charged chiral multiplets located at a boundary (see e.g.~\cite{Mirabelli:1997aj} for an explicit Lagrangian). This can for instance be an MSSM matter- or Higgs-sector. The inclusion of their effects in the Casimir energy is straightforward: The contribution of brane fields to the running of the gauge coupling is the usual one of 4d gauge theories. For fields at a brane (of an $ \orbi $) where the gauge symmetry is $H = H_1 \otimes \ldots \otimes H_n $ this yields\footnote{In our conventions, the $\beta$-function coefficient for fields charged under a gauge group $ G $ is
$
{b_{G}= \frac 1 6 \left[ (-22) \ C_{\sG} + 4 \ (\textrm{\# of Weyl-fermions in rep. }\sR) \ C_{\sR} + 2 \ (\textrm{\# of complex scalars in rep. }\sR') \ C_{\sR'} \right]\,.}  
$ 
} 
 \be \label{eq:c_brH_chi}
c^{(br)}_{loc} = \frac{c_{\omega}}{2 \pi^3} \sum_{i=1}^{n} d_{\sH_i} b_{H_i} \,.
\ee
Here, $ b_{H_i} $ is the $\beta$-function coefficient for the unbroken subgroup $ H_i $. For the case of a brane with unbroken gauge symmetry this becomes $ c^{(br)}_{loc} = (c_{\omega}/2 \pi^3) d_{\sG} b_{G}  $. For $ \orbib $, the only difference is an extra factor of $ 16 $ due to the reduced length of the interval. Observe that $c^{(br)}_{loc}$ is always positive since the $\beta$-function coefficient of chiral multiplets is positive. Hence, in situations where the contribution of the bulk field content alone to $ c^{(br)} $ is not positive and large as it needs to be, brane-localized chiral multiplets can help to assure a perturbatively controlled radion effective potential.

Let us now apply our results to supersymmetric $ SU(5) $-GUTs on $ \orbib $ such as those which were proposed in Ref.~\cite{Hebecker:2001wq}  (see also~\cite{Kawamura:2000ev}). These models have an unbroken $ SU(5) $ brane as well as a brane where the gauge symmetry is broken to the Standard Model (SM) gauge group ${SU(3) \times SU(2) \times U(1) }$. The gauge sector resides in the bulk and the Higgs sector is located on the SM brane (this avoids the doublet-triplet splitting problem). In Ref.~\cite{Hebecker:2001wq}, the matter sector is assumed to be located either completely on the SM brane or  in the bulk. As explained in detail in Ref.~\cite{Hebecker:2001wq}, in the latter case each bulk matter family consists of \textit{two} copies of a $(\mathbf{10 + \bar 5})$ with opposite $Z'_2$-parities, such that there is a full MSSM matter family at the zero mode level. 

We generalize the models of Ref.~\cite{Hebecker:2001wq} by allowing for an arbitrary distribution of the MSSM matter to the bulk and the SM brane, which is our `micro-landscape'. For any of the three families, we allow the freedom to have, instead of a pair with opposite $Z'_2$-parities, just one copy of a  $ \mathbf{\bar 5=(\bar 3,1)_{1/3}\oplus(1,2)_{-1/2}} $ in the bulk. Depending on the $Z'_2$-parity of this  $ \mathbf{\bar 5}$, there is either a $D_{R}=\mathbf{(\bar 3,1)_{1/3}}  $ or an $L_L=\mathbf{(1,2)_{-1/2}} $ at the zero mode level. Analogously, for one HMP transforming as a $ \mathbf{10=(3,2)_{1/6}\oplus(\bar 3,1)_{-2/3}\oplus(1,1)_{1}} $, one has either a $Q_L= \mathbf{(3,2)_{1/6}} $ or a $U_R\oplus E_R= \mathbf{(\bar 3,1)_{-2/3}\oplus(1,1)_{1}} $  at the zero mode level. Let us in the following consider the most general situation where  the zero modes of HMPs lead to $ r $ generations of $ U_R\oplus E_R $, $ s $ generations of $ L_L $, $ t $ generations of $ Q_L $ and $ u $ generations of $ D_R $ in the bulk (where $ r,s,t,u\in\{0,1,2,3\} $). Consequently, the remaining  $ (3-r) $ $ U_R\oplus E_R $ generations, $ (3-s) $ $ L_L $ generations,  $ (3-t) $ $ Q_L $ generations and $ (3-u) $ $ D_R $ generations must be located at the SM brane. 

We now determine the Casimir energy for this situation. Applying Eqs.~(\ref{eq:v1_vmp_orbib}) and (\ref{eq:v1_hmp_orbib}) and adding $ c^{(1)}_{grav}=-64c_{\omega} $ from the supergravity multiplet, one finds
\be
\frac{c^{(1)}}{c_{\omega}}=-160-16 r -8 s+96 t +48 u\,,
\ee
where we used $c_{2 \omega} = 4 c_{\omega}+ \cO(\omega^4) $. Here and in the following we neglect $\cO(\omega^4) $ effects. The brane contribution to the Casimir energy which is induced by the VMPs and HMPs in the bulk is given by Eqs.~(\ref{eq:c_brG_vmp}), (\ref{eq:c_brH_vmp}), (\ref{eq:c_brG_hmp}) and (\ref{eq:c_brH_hmp}).\footnote{To
evaluate Eq.~(\ref{eq:c_brG_hmp}) one needs $ C_2(\mathbf{\bar 5})=12/5 $ and $ C_2(\mathbf{10})=18/5 $. To evaluate Eq.~(\ref{eq:c_brH_hmp}) one needs
\be
\sum_{a, \bar \imath} \eta^a (T^a_{\mathbf{\bar 5}}T^a_{\mathbf{\bar 5}})_{\bar \imath \bar \imath}=-6/5\,, \qquad
\sum_{a, \hat \imath} \eta^a (T^a_{\mathbf{\bar 5}}T^a_{\mathbf{\bar 5}})_{\hat \imath \hat \imath}= 6/5\,, \nn
\ee
where $\bar \imath$ is an $L_L$-index and $\hat \imath$ a $D_{R} $-index, as well as
\be
\sum_{a, \bar \imath} \eta^a (T^a_{\mathbf{10}}T^a_{\mathbf{10}})_{\bar \imath \bar \imath}=-18/5\,, \qquad
\sum_{a, \hat \imath} \eta^a (T^a_{\mathbf{10}}T^a_{\mathbf{10}})_{\hat \imath \hat \imath}=18/5\,,  \nn
\ee
where $\bar \imath$ is a $U_R\oplus E_R$-index and $\hat \imath$ a $Q_L$-index.
}
Adding all terms, one finds after some algebra that
\be
\frac{c^{(br)}_{hmp+vmp}}{c_{\omega}}=-\frac{12}{5\pi^3}\left(50+27 r +9 s-57 t -19 u \right) \,.
\ee 
On the other hand, using Eq.~(\ref{eq:c_brH_chi}), the brane effect induced by the matter and Higgs fields on the brane is found to be
\be
\frac{c^{(br)}_{loc}}{c_{\omega}}=\frac{24}{5\pi^3}\left(126-9 r -3 s -21 t -7 u\right)\,,
\ee
so that the total brane coefficient is
\be
\frac{c^{(br)}}{c_{\omega}}=\frac{1}{\pi^3}\left(2424/5-108 r -36 s +36 t +12 u\right) \,.
\ee

One can now determine the position of the minimum of $ V(R) $ in units of $ g^2 $ or equivalently in units of $ 1/M  \simeq g^2 N / (24 \pi^3)$, for all choices of $ r, s, t,  u $. Note that the condition $ c^{(1)}<0  $ for the existence of a minimum is not satisfied for many choices of $ r, s, t,  u $, while  $c^{(br)}>0  $ is always satisfied. If a minimum exists, one easily finds that it is given by
\be
M R=\gamma \, W_{-1}\left( \sqrt[5]{e}/\gamma \right) \qquad \textrm{where} \qquad \gamma\equiv \frac{5}{4}\frac{24 \pi^3}{N}\frac{c^{(br)}}{c^{(1)}} \,.
\ee
Here, $ W_{-1}(x) $ is the Lambert $W$-function, which is defined as the inverse function of $ x\, e^x $. More precisely, since the Lambert $W$-function is double-valued on $(-1/e, 0)$,  $ W_{-1}(x) $ denotes the branch which satisfies $ W_{-1}(x)\le -1 $ for $-1/e\le x<0  $ (see e.g. \cite{lambert}). By scanning all $ 4^4=256 $ possible choices of $ r, s, t,  u $, we find that $ V(R) $ has a minimum at large radius, say $ MR>10 $, for about a third of them. 

Since the minimum $ R $ is given in units of $ 1/M$ (or equivalently $ g^2 $, as explained in the Introduction), by stabilizing the radius we determine the value of the 4d gauge coupling $g_4^2=g^2/(2 \pi R)$ at the scale $M_c$. The phenomenological value is $ \alpha(M_c) = g_4^2/4 \pi \simeq 1/25 $. For 12 choices of $ r, s, t,  u $, we find values for $ \alpha^{-1}(M_c)$ in the interval $20 \cdots 30$ (we give rounded values):
\begin{displaymath}
\begin{array}{|c||c|c|c|c|c|c|c|c|c|c|c|c|}
\hline
r&1&1&2&2&0&2&1&1&3&0&3&2\\
\hline
s&1&3&2&0&3&2&0&2&1&2&1&1\\
\hline
t&0&0&0&0&0&1&0&0&0&0&1&0\\
\hline
u&0&1&2&1&0&0&0&1&3&0&1&2\\
\hline\hline
\alpha^{-1}&20&20&22&22&23&24&25&26&26&28&29&30\\
\hline
\end{array}
\end{displaymath}
For some orbifold GUT models (albeit not the most simple ones) we predict a realistic size for the unified gauge coupling! 

As a comparison, for the most simple model where all matter is located at the SM brane (i.e.~${ r=s=t=u=0 }$) we find $\alpha\simeq1/40 $. This is too small. One should keep in mind however that the correction due to unknown nonvanishing but not unnaturally large coefficients of the brane-localized operators at the scale $ M $ is expected to be roughly of the order $ 1/\ln(MR)\sim 25\% $.

\section{Uplifting to a small cosmological constant} \label{sec:uplifting}

The Casimir energy we find gives a negative contribution of the order 
$\omega^2/R^4 \sim m_{1/2}^2 M_c^2$ to the vacuum energy. In order for the 
theory to be potentially realistic, there need to be other effects 
canceling this negative contribution, such that a tiny positive $\Lambda_4$
is obtained. This will certainly involve fine-tuning, otherwise we would 
have solved the cosmological constant problem.\footnote{
Given 
our string-theoretic motivation, which relies mainly on the recent 
progress in heterotic orbifold model building, it is tempting to ascribe 
the required fine tuning to the multitude of vacua in the string theory 
landscape. The problem with this argument is our insufficient 
understanding of the heterotic landscape, which at present does not allow
us to find a sufficiently large and dense discretuum within the 
relevant orbifold constructions (unlike the type IIB case, where 
the situation is more promising from the perspective of the cosmological 
constant). For the purpose of this paper, we take the optimistic attitude 
that either such a heterotic landscape will be found or the cosmological
constant problem will be solved in some other way.
}
In this section, we briefly discuss how such an `uplifting' could be realized. 
We also check that our proposals are consistent with the assumptions of our 
Casimir energy calculation, namely a (sufficiently) flat 5d background and 
Scherk-Schwarz SUSY breaking.

Note that there are no loop-contributions to $\Lambda_4$ coming from fields 
other than the radion. This is easily seen in the language of 4d supergravity: 
Our Casimir energy calculation is equivalent to the calculation of a loop 
correction to the no-scale K\"ahler potential of the radion. In the presence 
of a constant brane-localized superpotential $W_0$, which is related to the 
Scherk-Schwarz parameter $\omega$ by
\be
\omega \sim \frac{|W_0|}{M_{P,5}^3} \,,
\ee
this correction turns into a potential energy. Loop effects will generically 
also correct the K\"ahler potential of matter and Higgs fields, which is 
canonical at tree level. Since these fields do not develop a VEV, their 
K\"ahler corrections do not induce a contribution to the vacuum energy. 
Furthermore, there are no perturbative corrections to the superpotential. 
Thus, the negative vacuum energy which we find at the minimum of the radion 
effective potential has to be taken seriously and some compensating effect
is required.

\subsection{Uplifting by small warping and a brane-superpotential}\label{bsp}

Let us allow for a 5d cosmological constant, which of course has to be 
small enough not to affect our flat-space Casimir energy calculation. In 
other words, we assume that we are dealing with a supersymmetric 
Randall-Sundrum model~\cite{Randall:1999ee}, but with very weak warping. The 
drawback of this proposal is that we do not know how such a warping could 
arise from the heterotic orbifold perspective. One may hope that 
it can be realized, e.g., by fluxes in the five compact dimensions, 
which would make it discretely tunable. In any case, what follows 
should be consistent from the point of view of 5d supergravity coupled 
to gauge fields and charged matter.

In the presence of a constant superpotential $W_0$ at the IR brane, the 
corresponding 4d theory is defined by~\cite{Luty:2000ec}
\be
\Omega = \frac{3 M_{P,5}^3}{k} \left( e^{-k (T + \bar T)}-1 \right) \qquad
\mbox{and}\qquad W=W_0\,e^{-3kT}\,.\label{omw}
\ee
The AdS curvature scale $k$ is related to the 5d cosmological constant 
by $\Lambda_5=-6k^2M_{P,5}^3$. Working on $S^1/Z_2$, the $T=\pi R+\cdots$ 
is the radion superfield, while $\Omega$ and $W$ are the `superspace kinetic 
function' and the superpotential of 4d supergravity. A K\"ahler-Weyl 
rescaling brings them to the equivalent form 
\be
\Omega = \frac{3 M_{P,5}^3}{k} \left( 1- e^{k (T + \bar T)} \right) \equiv 
-3 M_{P,5}^3 (T + \bar T) + \Delta \Omega \qquad\mbox{and}\qquad W=W_0\,.
\ee

For weak warping, $k (T + \bar T) \ll 1 $, we have
\be
\Delta \Omega \simeq - \frac 3 2 M_{P,5}^3 k (T + \bar T)^2 \,,\label{wic}
\ee
which can then be treated as a small correction to the basic no-scale 
structure of the model. This puts us into the setting of `Almost no-scale
supergravity' of Luty and Okada~\cite{Luty:2002hj}, where the corresponding 
correction to the Brans-Dicke-frame scalar potential, 
\be
\delta V = - \frac{|W_0|^2}{M_{P,5}^6} (\Delta \Omega)_{T \bar T}=3 \, k 
\frac{|W_0|^2}{M_{P,5}^3} \,,\label{delV}
\ee
has been given.\footnote{
Beware 
of a typo in Eq.~(6) of the arXiv-version of~\cite{Luty:2002hj}.
}
One immediately sees that, in order for $ \delta V $ to cancel the negative 
Casimir energy, one needs a warping of the order 
\be
k (T + \bar T) \sim \left( \frac{M_c}{M_{P,5}} \right)^3 \,.
\ee
Thus, the warping required for the uplifting is indeed small whenever the 
stabilization radius is large in units of the 5d Planck scale (this is anyway 
necessary for gravity to be perturbative at the compactification scale). We 
conclude that our flat space calculation remains justified in spite of the 
fact that we are really dealing with a Randall-Sundrum type model. 

We finally note that our complete stabilization and uplifting proposal can 
be formulated within the framework of~\cite{Luty:2002hj}: Our two-loop 
Casimir energy can be reinterpreted as a K\"ahler correction with the structure
\be
\Delta\Omega_{\rm Casimir} \sim \frac{1}{(T+\bar{T})^2}+\frac{g^2}
{(T+\bar{T})^3}\ln(M(T+\bar{T}))\,.
\ee
Adding this to the warping-induced correction of Eq.~(\ref{wic}), we find that 
Eq.~(\ref{delV}) generates a scalar potential the minimum of which can be 
tuned to zero by adjusting the ratios of $k$, $M_{P,5}$ and $g$.

\subsection{Uplifting in a detuned Randall-Sundrum model}\label{det}
The uplifting proposal of the last subsection can be reformulated in terms 
of the `supersymmetric detuned Randall-Sundrum model' of Bagger and 
Belyaev~\cite{Bagger:2002rw}. According to~\cite{Bagger:2002rw}, the UV and 
IR brane tensions $\Lambda_0$ and $\Lambda_\pi$, which are normally given by 
\be 
\Lambda_0=-\Lambda_\pi=\sqrt{-6 \Lambda_5 M_{P,5}^3}\equiv\Lambda\,,
\ee
can take arbitrary values in a consistent 5d supergravity model, as long as
they obey the constraint $|\Lambda_{0,\pi}|\leq \Lambda$. Thus, it is natural 
to attempt to uplift our previous stabilized flat 5d model by allowing for 
a small warping together with a small detuning of the IR-brane tension,
\be
0<\Lambda_\pi+\Lambda\ll\Lambda\,.
\ee
We keep $\Lambda_0=\Lambda$ for simplicity. 

The naive expectation is that, as long as warping is small, this will give 
a constant and positive contribution to the radion effective potential in the 
Brans-Dicke frame. Note that this is not inconsistent with general theorems 
concerning the possible vacuum states of supergravity theories: The detuned 
Randall-Sundrum model has an AdS$_4$ ground state at a certain radius. Since
we stabilize the radius by the Casimir energy at a different value, we are 
actually forcing the theory into a metastable state with a tiny positive 
$\Lambda_4$.

To confirm the above expectation, we utilize the 4d supergravity 
description of the detuned Randall-Sundrum model derived in~\cite{
Bagger:2003dy}. The K\"ahler potential $K$ and the superpotential $W$ are 
explicitly given in Eqs.~(6.1) of~\cite{Bagger:2003dy}. We first rewrite
$K$ in terms of $\Omega=-3\exp(-K/3)$ and perform a (constant) K\"ahler-Weyl
transformation bringing $\Omega$ to the form given in Eq.~(\ref{omw}). We 
then work out $W$ in the limit $\Lambda_0\to\Lambda$, making also 
use of the relation $\Lambda_\pi+\Lambda\ll\Lambda$. The functional form 
agrees with our Eq.~(\ref{omw}) and we determine
\be
|W_0|=\sqrt{2}M_{P,5}^3\sqrt{1+\Lambda_\pi/\Lambda}\,.
\ee
Calculating $\delta V$ according to Eq.~(\ref{delV}), we find
\be
\delta V = \Lambda_\pi+\Lambda\,,\label{upp}
\ee
as expected (see also~\cite{Falkowski:2005fm}). We conclude that the uplifting 
proposals of Sect.~\ref{bsp} and of the present section are equivalent. The 
underlying technical result is the superfield formulation of the detuned 
Randall-Sundrum model of~\cite{Bagger:2003dy}: It implies that including a 
constant IR-brane-localized superpotential in a supersymmetric Randall-Sundrum 
model is equivalent to a weak detuning of the IR brane tension. In both cases, 
a non-zero scalar potential is induced. This potential is positive and 
approximately constant at small values of the radion (i.e.~for a warp 
factor close to one). 

We note that the 4d superfield description of the detuned Randall-Sundrum model
has also been considered, e.g., in~\cite{Falkowski:2005fm} (independently 
of~\cite{Bagger:2003dy}) as well as in~\cite{Katz:2005wp} and~\cite{
Correia:2006vf}. In particular, one-loop corrections to the radion potential 
have been analyzed in~\cite{Falkowski:2005fm,Katz:2005wp}. Our `uplifting'
proposal of Sects.~\ref{bsp} and~\ref{det} differs in that we use a two-loop 
effective potential to stabilize the radion at a very small (from the 
Randall-Sundrum model perspective) value. The warping is then irrelevant for 
the loop calculation and its only effect is to provide, in its interplay with 
a small detuning, an approximately constant uplifting contribution. For 
related applications of the detuning in supersymmetric Randall-Sundrum models
see, e.g.,~\cite{vonGersdorff:2003rq,Maru:2006ji}.

\subsection{$F$-term uplifting}
Alternatively, one may insist that $ \Lambda_5 $ is exactly zero. In that 
case, Scherk-Schwarz breaking predicts a negative vacuum energy and 
there needs to be another source of SUSY breaking, in the spirit of 
`$F$-term uplifting' (see e.g.~\cite{GomezReino:2006dk}). Following 
again~\cite{Luty:2002hj}, we assume that via some unspecified dynamics 
there arises an $F$-term VEV of a brane-localized singlet $S$ of the right 
order of magnitude to obtain a small cosmological constant: $F_S\sim 
\omega/R^2$. This can potentially affect the SUSY-breaking mass 
splitting of bulk fields and hence our Casimir energy calculation. 

In the spirit 
of gaugino mediation, $S$ may couple to the gauge-kinetic term via a 
brane-localized higher-dimension operator, suppressed by the fundamental 
scale $M$. The induced gaugino masses are of the order~\cite{Kaplan:1999ac}
\be
m_{1/2}\sim \frac{F_S}{M^2 R} \sim \frac{\omega}{M^2 R^3}\,.\label{gm}
\ee
Similarly, $S$ may couple to the kinetic term of bulk hypermultiplets via a
higher-dimension brane-localized operator. This induces scalar masses of the 
same order of magnitude as the gaugino masses given in Eq.~(\ref{gm}), 
$m_0\sim m_{1/2}$. 

By contrast, the `radion-mediated contribution' to the gaugino masses and to
the scalar masses of bulk multiplets is of the order 
\be
m_0\sim m_{1/2}\sim \frac{\omega}{R}\,,
\ee
so that in comparison the effect of $F_S$ is suppressed by $1/(M R)^2$. We 
conclude that our Casimir energy calculation and the corresponding 
stabilization mechanism remain under quantitative control in the presence 
of brane-localized $F$-term uplifting. One may nevertheless feel that the 
uplifting mechanism of Sects.~\ref{bsp} and~\ref{det} is more elegant since 
it does not require an extra SUSY breaking sector.

\section{Conclusions}

We have analyzed Casimir stabilization of supersymmetric gauge 
theories on 5d orbifolds with Scherk-Schwarz SUSY breaking and gauge 
symmetry breaking by boundary conditions. Depending on field content and 
symmetries of the 5d theory, a minimum in the Casimir energy (as a function 
of the radius~$ R $) can arise from the interplay of the one-loop and two-loop 
contribution. The dominant two-loop effect comes from logarithmically 
divergent operators localized at the boundaries. We rely only on the 
log-enhanced part of the coefficients of these operators. Our results are 
therefore independent of the UV completion as long as the compactification 
scale $ 1/R $ is much smaller than the cutoff scale $ \Lambda $. Then, the 
Casimir energy~-- including the one-loop term~-- is given by 
(see Eq.~(\ref{eq:V(R)_orbi}))
\be 
V(R)\,=\,\frac{1}{R^{4}}\left( c^{(1)}+c^{(br)}\,\ln(\Lambda R)\, 
\frac{g^{2}}{R} \right) \,.
\ee

We provide general formulae for the one-loop coefficient $c^{(1)}$ and for 
the coefficient $ c^{(br)} $ of the brane-induced two-loop effect. The 
coefficient $c^{(1)}$ is well-known for the case of unbroken gauge symmetry 
(Eq.~(\ref{eq:c1bulk})). It is unchanged for $ \orbi $ with broken gauge 
symmetry. For $ \orbib $ with gauge symmetry broken at one of the branes, 
it is given by Eq.~(\ref{eq:v1_vmp_orbib}) for vector- and by 
Eq.~(\ref{eq:v1_hmp_orbib}) for hypermultiplets. 

For $\orbi$, the coefficient $c^{(br)}$ is given by Eq.~(\ref{eq:c1br_total}) 
for unbroken gauge symmetry and by Eq.~(\ref{eq:orbi-broken-result}) for 
gauge symmetry breaking by boundary conditions. In the most relevant case 
of $\orbib$ with gauge symmetry breaking at one of the boundaries, $c^{(br)}$
is given by Eqs.~(\ref{eq:c_brG_vmp}), (\ref{eq:c_brH_vmp}) for 
vector multiplets and by Eqs.~(\ref{eq:c_brG_hmp}), (\ref{eq:c_brH_hmp}) for 
hypermultiplets. We were able to obtain these coefficients without explicit 
new loop calculations, relying only on the $Z_2$ parity transformation 
properties of the fields, supersymmetry and the group theoretic structure 
of the model. 

We have applied the above formulae to $SU(5)$ orbifold GUT models on 
$ \orbib $, with gauge-symmetry breaking to the Standard Model at one of the 
boundaries. For simplicity, we have assumed that the cutoff $\Lambda$ takes 
its highest possible value -- the strong coupling scale ${ M\simeq 24 
\pi^3/(5 g^2) }$ of the 5d gauge theory. Furthermore, we have focused on 
scenarios where the Higgs sector is located at the Standard Model brane,
allowing the matter sector to be distributed between bulk and Standard Model 
brane in various ways. The resulting 256 possibilities form our 
`micro-landscape'. Of course, using our formulae, one could also consider 
models which, for instance, have Higgs fields in the bulk instead of on the 
brane and/or have additional vector-like matter. Similarly, $SO(10)$ or 
other unified gauge groups could be studied. Our analysis is clearly 
incomplete from this perspective. Nevertheless, restricting ourselves to the 
simple class of GUT models defined above, we find a minimum in the effective 
potential at large $ MR $ for many of the possible matter distributions. 
The position of the minimum determines the 4d gauge coupling at the 
compactification scale. For several of the models, we find results which are 
in agreement with the phenomenological value of ${ \alpha_{GUT}\simeq1/25 }$. 
Thus, two-loop Casimir stabilization offers a simple `explanation' (within 
our modest realization of the landscape paradigm) for the appearance of a 
relatively large 5th dimension and a correspondingly small gauge coupling
at the GUT scale.

We have finally discussed two mechanisms for uplifting our AdS$_4$ vacua 
without affecting the underlying stabilization mechanism: One possibility is 
to allow for a small warping, i.e. a small negative 5d cosmological constant. 
In combination with a brane localized constant superpotential, this 
induces the required positive contribution to the scalar potential. 
Alternatively, one can include an extra SUSY-breaking sector on one of the 
branes (independently of the radion, which is our dominant source of
SUSY breaking). Such a brane-localized $F$-term can provide the required 
uplift. It would be important to find a string-theoretic realization of 
these uplifting mechanisms and, more generally, to work out in more detail 
to which extent our stabilization proposal is consistent with a full-fledged 
underlying heterotic string construction.

\noindent
\textbf{Acknowledgments:} We would like to thank Gero von Gersdorff, Tony Gherghetta and Marco Serone for useful comments and discussions. CG is supported by the German Science Foundation (DFG) under the Collaborative Research Center (SFB) 676.

\begin{appendix}

\renewcommand{\theequation}{A.\arabic{equation}}
\setcounter{equation}{0}  

\section*{Appendix}

We find that some of our results are not consistent with Ref.~\cite{Cheng:2002iz} where the one-loop mass shifts of the KK-modes due to brane-localized effective operators in a 5d Universal-Extra-Dimensions scenario are calculated. One should be able to obtain our formulae for $ c^{(br)} $ by inserting the shifted KK masses which are calculated in Ref.~\cite{Cheng:2002iz} (they discuss only unbroken gauge symmetry however) into Eq.~(\ref{eq:Vbr}). We believe there is an error in the result for the mass corrections of fermions due to scalars in the loop (see Eq.~(39) and (B5) in Ref.~\cite{Cheng:2002iz}). In those equations, the number `3' which appears twice should each time be replaced by a `-1'. As a side remark, this implies that the KK masses of the third generation left-handed quark doublet and right-handed top receive a positive contribution due to the Yukawa coupling instead of a negative one (cf. Eq.~(45) in Ref.~\cite{Cheng:2002iz}). 

We briefly outline the computation we performed as a check (following~\cite{Georgi:2000ks}). Consider the Yukawa theory given by the action\footnote{Our conventions for the 5d gamma-matrices are $\Gamma^M\equiv(\gamma^{\mu},i \gamma^5)$ where the $\gamma^{\mu}$ are generators of the 4d Clifford algebra and $\gamma^5\equiv\gamma_5\equiv i \gamma^0\gamma^1\gamma^2\gamma^3$.}
\be \label{eq:treelag}
\int\limits \!\! d^{4}x \!\!\! \int\limits_{-\pi R}^{\pi R} \!\!\! dy \sqrt{- \det (g_{M N})} \left( \frac{1}{2}M_{P,5}^{3} \mathcal{R}_{5} +   \frac{1}{2}\partial_{M}\phi \ \partial^{M}\phi + i \bar{\psi} \Gamma^{M} \partial_{M} \psi - h \bar{\psi} \psi \phi \right) .
\ee
Consistency requires the parities to be $\phi(-y)=- \phi(y)$ and $\psi(-y)=Z_{\psi} \gamma_{5} \psi(y)$ where $Z_{\psi}\in\{\pm 1\}$. The propagators for the 4d KK modes are
\bea
\langle \phi^{(n)}(p)\phi^{(n')}(p) \rangle&=& \frac{i}{2} \frac{1}{p^{2}-(\frac{n}{R})^2} ( \de_{n,n'} - \de_{n, -n'} ) \nonumber 
\\
\langle \psi^{(n)}(p)\bar{\psi}^{(n')}(p) \rangle &=& \frac{i}{2} \frac{\slashed{p}+i\gamma_5\frac{n}{R}}{p^{2}-(\frac{n}{R})^2} ( \de_{n,n'} - Z_{\psi} \gamma_5 \ \de_{n, -n'} ).
\eea
The fermion selfenergy (times $-i$) due to a scalar in the loop then is
\bea 
&&\begin{picture}(140,100)(0,30)
\unitlength=1.2 pt
\SetScale{1.2}
\SetWidth{0.7}      
\scriptsize    
\ArrowLine(5.0,50.0)(40.0,50.0)
\ArrowLine(40.0,50.0)(100.0,50.0)
\ArrowLine(100.0,50.0)(135.0,50.0)
\DashCArc(70,50)(30,0,180){3}
\Text(70,40)[]{$k$}
\Text(50,40)[]{$\frac{m'}{R}$}
\Text(90,40)[]{$\frac{m}{R}$}
\Text(70,90)[]{$p-k$}
\Text(30,70)[]{$\frac{n'-m'}{R}$}
\Text(110,70)[]{$\frac{n-m}{R}$}
\Text(13,58)[]{$p,\frac{n'}{R}$}
\Text(127,58)[]{$p,\frac{n}{R}$}
\Text(95,85)[]{($\phi$)}
\Text(10,40)[]{($\psi$)}
\Text(0,50)[]{$i'$}
\Text(140,50)[]{$i$}
\end{picture}
\nonumber\\
&=&\frac{h^{2}}{8 \pi R} \sum_{m} I(\slashed{p},m) \big( \underbrace{ \de_{n,n'} + Z_{\psi}\de_{n,-n'} \g_{5} }_{\rightarrow bulk}  \underbrace{- \de_{2m,(n+n')} - Z_{\psi} \de_{2m,(n-n')}\g_{5}}_{\rightarrow boundary} \big)
\eea
where
\bea
I(\slashed{p},m) &\equiv& \mu^{2 \e}\dke \frac{\slashed{k}+i\gamma_{5}\frac{m}{R}}{(k^{2}-(\frac{m}{R})^{2})[(p-k)^{2}-(\frac{n-m}{R})^{2}]} \nonumber
\\
&\longrightarrow& -i\frac{1}{16 \pi^{2}}\ln \frac{\Lambda^{2}}{\mu^{2}} \lb \frac{\slashed{p}}{2} + i\gamma_{5}\frac{m}{R}\rb
\eea
and the arrow means taking the divergent part with cutoff $\Lambda$ and letting $\epsilon \rightarrow 0$.
We can then write the selfenergy contribution due to the boundary as
\be \label{eq:selfener}
\sum_{m} \left( a_{1} \frac{\slashed{p}}{2} + a_{2} i\gamma_{5}\frac{m}{R} \right) (\de_{2m,(n+n')} + Z_{\psi} \gamma_5 \de_{2m,(n-n')})
\ee
where
\be
a_{1} = -\dfrac{1}{64 \pi^{2}} \frac{h^2}{2 \pi R} \ln  \frac{\Lambda^{2}}{\mu^{2}}  = a_{2}.
\ee
The effective Lagrangian which follows from Eq.~(\ref{eq:selfener}) by Fourier transformation is
\be
\delta\cL_{5}=\frac{\delta(y)+\delta(y-\pi R)}{2} (2 \pi R) \Big\{ a_1 i \bar{\psi}_{+} \ds \psi_{+} + Z_{\psi} a_2  \big[ (\partial_{5} \bar{\psi}_{-}) \psi_{+} + \bar{\psi}_{+} (\partial_{5} \psi_{-}) \big] \Big\}
\ee
where $\psi_{\pm}=\frac 1 2 (1 \pm Z_{\psi} \g_5) \psi$. Expanding this into KK modes and comparing it to the tree-level Lagrangian~(\ref{eq:treelag}), one finds that the $n$th KK mode receives a mass shift
\be
\delta m_{n}=m_{n} (2 a_2 - a_1) = - m_{n} \dfrac{1}{64 \pi^{2}} \frac{h^2}{2 \pi R} \ln  \frac{\Lambda^{2}}{\mu^{2}}.
\ee
This disagrees with Eq.~(39) in Ref.~\cite{Cheng:2002iz}. Note that, apart from this direct calculation, one also sees that there is an inconsistency among Eqs.~(38) and (39) in Ref.~\cite{Cheng:2002iz} because, applied to a supersymmetric theory, they imply that KK gauge bosons and KK gauginos would not have equal masses.

\end{appendix}

\end{document}